\documentclass[10pt,conference]{IEEEtran}
\usepackage{graphicx}
\usepackage{subcaption}
\usepackage{float}
\usepackage{booktabs}
\usepackage{cite}
\usepackage{amsmath,amssymb,amsfonts}

\usepackage[colorlinks]{hyperref}
\usepackage{xcolor}          
\def\tmp#1#2#3{%
  \definecolor{Hy#1color}{#2}{#3}%
  \hypersetup{#1color=Hy#1color}}
\tmp{link}{HTML}{800006}
\tmp{cite}{HTML}{2E7E2A}
\tmp{file}{HTML}{131877}
\tmp{url} {HTML}{8A0087}
\tmp{menu}{HTML}{727500}
\tmp{run} {HTML}{137776}
\def\tmp#1#2{%
  \colorlet{Hy#1bordercolor}{Hy#1color#2}%
  \hypersetup{#1bordercolor=Hy#1bordercolor}}
\tmp{link}{!60!white}
\tmp{cite}{!60!white}
\tmp{file}{!60!white}
\tmp{url} {!60!white}
\tmp{menu}{!60!white}
\tmp{run} {!60!white}

\usepackage{mathtools}
\DeclarePairedDelimiter{\ceil}{\lceil}{\rceil}

\title{Comparing Algorithms for Loading Classical Datasets into Quantum Memory}

\begin{document}

\author{\IEEEauthorblockN{Andriy Miranskyy}
\IEEEauthorblockA{\textit{Dept. of Computer Science} \\
\textit{Toronto Metropolitan University} \\
Toronto, Canada \\
avm@torontomu.ca}
\and
\IEEEauthorblockN{Mushahid Khan}
\IEEEauthorblockA{\textit{Dept. of Electrical and Computer Engineering} \\
\textit{University of British Columbia} \\
Vancouver, Canada \\
mkhan103@student.ubc.ca}
\and
\IEEEauthorblockN{Udson Mendes}
\IEEEauthorblockA{\textit{Quantum Technologies Team} \\
\textit{CMC Microsystems} \\
Sherbrooke, Canada \\
udson.mendes@cmc.ca}

}

\maketitle

\begin{abstract}
Quantum computers are gaining importance in various applications like quantum machine learning and quantum signal processing. These applications face significant challenges in loading classical datasets into quantum memory. With numerous algorithms available and multiple quality attributes to consider, comparing data loading methods is complex.

Our objective is to compare (in a structured manner) various algorithms for loading classical datasets into quantum memory (by converting statevectors to circuits).

We evaluate state preparation algorithms based on five key attributes: circuit depth, qubit count, classical runtime, statevector representation (dense or sparse), and circuit alterability. We use the Pareto set as a multi-objective optimization tool to identify algorithms with the best combination of properties. To improve comprehension and speed up comparisons, we also visually compare three metrics (namely, circuit depth, qubit count, and classical runtime).

We compare seven algorithms for dense statevector conversion and six for sparse statevector conversion. Our analysis reduces the initial set of algorithms to two dense and two sparse groups, highlighting inherent trade-offs.

This comparison methodology offers a structured approach for selecting algorithms based on specific needs. Researchers and practitioners can use it to help select data-loading algorithms for various quantum computing tasks.

\end{abstract}

\section{Introduction}\label{sec:intro}
Quantum computers are rapidly evolving, demonstrating utility even before reaching the stage of fault-tolerant devices~\cite{kim2023evidence,murillo2024challenges}. 

Loading classical datasets into a quantum computer is a critical step in many quantum computing use cases, such as solving systems of equations, quantum machine learning, and quantum signal processing~\cite{zhang2021low}. Once the dataset has been loaded, a quantum algorithm can be applied to analyze it. Typically, this involves encoding  $N$  data points (which can be represented by $N$ complex or real numbers) into the states of  $n = \ceil{\log_2(N)}$  qubits.

Using a classical computer, we can reduce this task to the following: $N$  data points are converted to $N$  bit strings, which are then mapped to integers. These integers become indexes of nonzero elements of a statevector. The statevector is then rendered in the memory of a classical computer, where an algorithm converts it to a quantum circuit. The conversion to a circuit of the entire statevector is exponentially complex: the lower bound for state preparation with approximation to within a distance $\epsilon$ is $\Omega(2^n \log(1/\epsilon)/\log(n))$~\cite[see Sec.~4.5.4 and Eq.~4.85]{nielsen_chuang_2010}.

\subsection{Quality attributes}
Statevectors can be converted to circuits using a variety of algorithms. How do we select the most suitable algorithm for our needs? In order to help us make our selection, let us examine five quality attributes of these algorithms.

The three quality attributes of a state preparation algorithm~---~circuit depth, qubit count, and classical runtime~---~cannot be optimized simultaneously due to inherent trade-offs in computational resources and constraints. Informally, we can think of it as follows:

\begin{enumerate}
    \item \textit{Circuit Depth}: Reducing the depth of the circuit reduces\footnote{Once a circuit is obtained via a specific algorithm designed to load a classical dataset, the circuit can be further optimized using generic techniques incorporated into quantum computer compilers and optimizers, such as reinforcement learning-based methods~\cite{kremer2024practical}. This optimization process is made easier with a well-designed initial circuit.} the quantum runtime and mitigates decoherence (quantum errors resulting from noise over time). However, achieving a shallow circuit requires more qubits and increased complexity in preparing the initial state, leading to a longer classical runtime.
    \item \textit{Qubit Count}: Minimizing the qubit count is desirable because quantum hardware is still limited in the number of qubits. However, reducing qubit count often results in a deeper circuit since fewer qubits must perform more operations. This can increase both circuit depth and classical runtime.
    \item \textit{Classical Runtime}: Decreasing the classical runtime, or the time spent on a classical computer preparing the quantum circuit, can be achieved by simplifying the data encoding process. However, this simplification can result in a deeper circuit (more quantum operations) or an increase in the number of qubits needed to represent the data.
\end{enumerate}

The fourth quality attribute~---~statevector representation (dense or sparse\footnote{It is often possible to reduce classical datasets from machine learning and similar fields to sparse statevectors.})~---~stems from the exponential growth of states with the number of qubits, which can exhaust a classical computer's memory. A dense algorithm loads the entire statevector into memory, while a sparse algorithm loads only those elements that are not zero. A dense statevector algorithm must encode $N=2^n$ objects, while a sparse statevector algorithm~--- only $N=r$ objects, where $r$ is the number of non-zero elements in the statevector.
\begin{enumerate}
    \setcounter{enumi}{3}
    \item \textit{Statevector Representation}: Using a sparse statevector representation reduces memory footprint on a classical computer (in comparison with the dense representation), but leads to a longer classical runtime.
\end{enumerate}

Finally, let us examine the quality attributes from the perspective of online and offline use cases. Offline machine learning or signal processing involves accumulating data over time and training the entire model. In online use cases (e.g., pattern
mining or near-real-time anomaly detection~\cite{gomes2019machine,islam2020anomaly,islam2021anomaly}), data are acquired sequentially, and the predictor is updated with new data as it comes in, sometimes forgetting some of the older data. We \textbf{posit} that online use cases require the ability to dynamically reconstruct the circuit, adding or removing data points on the fly without recomputing the entire circuit from scratch. 
\begin{enumerate}
    \setcounter{enumi}{4}
    \item \textit{Alterable circuit}: Adding or removing data points to/from the circuit without recomputing the whole circuit. This capability will speed up classical runtime (essential for real-time applications using online algorithms). However, the circuit depth will also increase because the circuit cannot be optimized globally.
\end{enumerate}

As mentioned above, due to this interplay of constraints, it is impossible to achieve optimal levels of all quality attributes simultaneously. Rather, a balance must be struck according to the specific requirements and limitations of each quantum computing task.

Let us explore how to compare algorithms.

\section{Crude algorithm comparison}
Ideally, we should implement all possible data loading algorithms, compare them (e.g., using the time-to-solution metric~\cite{pokharel2023demonstration}) on the data we need to load on a reference quantum computer, and choose the best algorithm. However, implementing all algorithms can be prohibitively expensive. Instead, we describe a crude comparison process (based on the five quality attributes mentioned above) that helps us to focus on a subset of algorithms. We can also add additional constraints and requirements (e.g., how many qubits we have and how much time we are given) as needed.

As a measure of comparison, we can use complexity metrics\footnote{It is important to note that comparing the exact number of, e.g., gates, would be more accurate than comparing the order of circuit depth. Some authors provide this information, but not all, making the complexity metric a more universal measure, although less accurate, as these formulas describe asymptotic behaviour rather than exact values.} provided by the algorithm authors. We will examine algorithms that use dense and sparse statevector representation independently, since the former depends only on $n$ and the latter~---~on $n$ and $r$.

We aim to demonstrate the variability of the algorithms' landscape and illustrate the comparison principles. Thus, the algorithms under study represent a sample rather than an exhaustive set. 

We need to resort to multi-objective optimization to compare algorithms (based on the order of circuit depth, classical runtime, and qubit count) and assess which algorithms have the best combined properties. We will compute a Pareto (non-dominiating) set\footnote{A point is in the Pareto set if: 1) it is not dominated by any other solution in the decision space; 2) there is no other solution that can improve one objective without sacrificing another, see~\cite{zitzler2008quality} for review.} using \texttt{paretoset} library v.1.2.3~\cite{paretoset}. We assume that each order value has equal weight for a given $n$ and $r$. The reader may adjust the weight according to their specific use case. Moreover, if necessary, the reader may choose a different multi-objective optimization method.

\begin{table*}[htb]
\caption{Quality attributes. A question mark indicates cases in which the authors do not explicitly report the value, and we approximate it to the best of our ability. We assume that $\log(\cdot)$ is a binary logarithm. Alterable circuit is discussed in Section~\ref{sec:alter}.}
\label{tbl:complexity}
\centering
\begin{tabular}{@{}llllll@{}}
\toprule
Algorithm Group               & Circuit Depth  & Classical Runtime & Qubit Count & Statevector Representation   \\ \midrule
Araujo'21~\cite{araujo2021divide}                & $O(\log(N)^2)$ &             $O(N)$            & $O(N)$      & Dense                                      \\
Unitary~\cite{mottonen2005transformation,plesch2011quantum} & $O(N)$                     & $O(N)$            & $O(n)$      & Dense                                      \\
Zhang'21a~\cite{zhang2021low}                & $O(n^2)$                   & $O(N^2)$          & $O(n)$      & Dense                                      \\
Zhang'21b~\cite{zhang2021low}                & $O(n^2)$                   & $O(n^2)$          & $O(N^2)$    & Dense                                      \\
Zhang'21c~\cite{zhang2021low}                & $O(n^2)$                   & $O(N^{1.52})$     & $O(N)$      & Dense                                      \\
Zhang'22a~\cite{zhang2022quantum}                & $\Theta(n)$                    & $O(N)$            & $O(N)$      & Dense                                      \\
\midrule
deVeras'22~\cite{deVeras2022double} & $O(nr)$?      & $O(nr + r \log(r))$  & $O(n)$              & Sparse  \\
Gleinig'21~\cite{gleinig2021efficient} & $O(nr)$        & $O(n r^2 \log(r))$ & $O(n)$              & Sparse   \\
NR-group~\cite{deVeras2020circuit,malvetti2021quantum,khan2022ep}         & $O(nr)$        & $O(nr)$            & $O(n)$              & Sparse   \\
Zhang'22b~\cite{zhang2022quantum}  & $\Theta(\log(nr))$        & $O(\log(nr))$      & $O(n+ n r \log(r))$ & Sparse  \\
\bottomrule        
\end{tabular}%
\end{table*}

\subsection{Dense statevector representation}\label{sec:dense}
Table~\ref{tbl:complexity} gives the order of complexity for six groups of algorithms (seven algorithms in total) operating on dense statevectors. The algorithms are grouped by distinct values of the order of circuit depth, classical runtime, and qubit count.

\begin{figure*}[tb]
    \centering
    \begin{subfigure}[b]{0.32\textwidth}
        \centering
        \includegraphics[width=\textwidth]{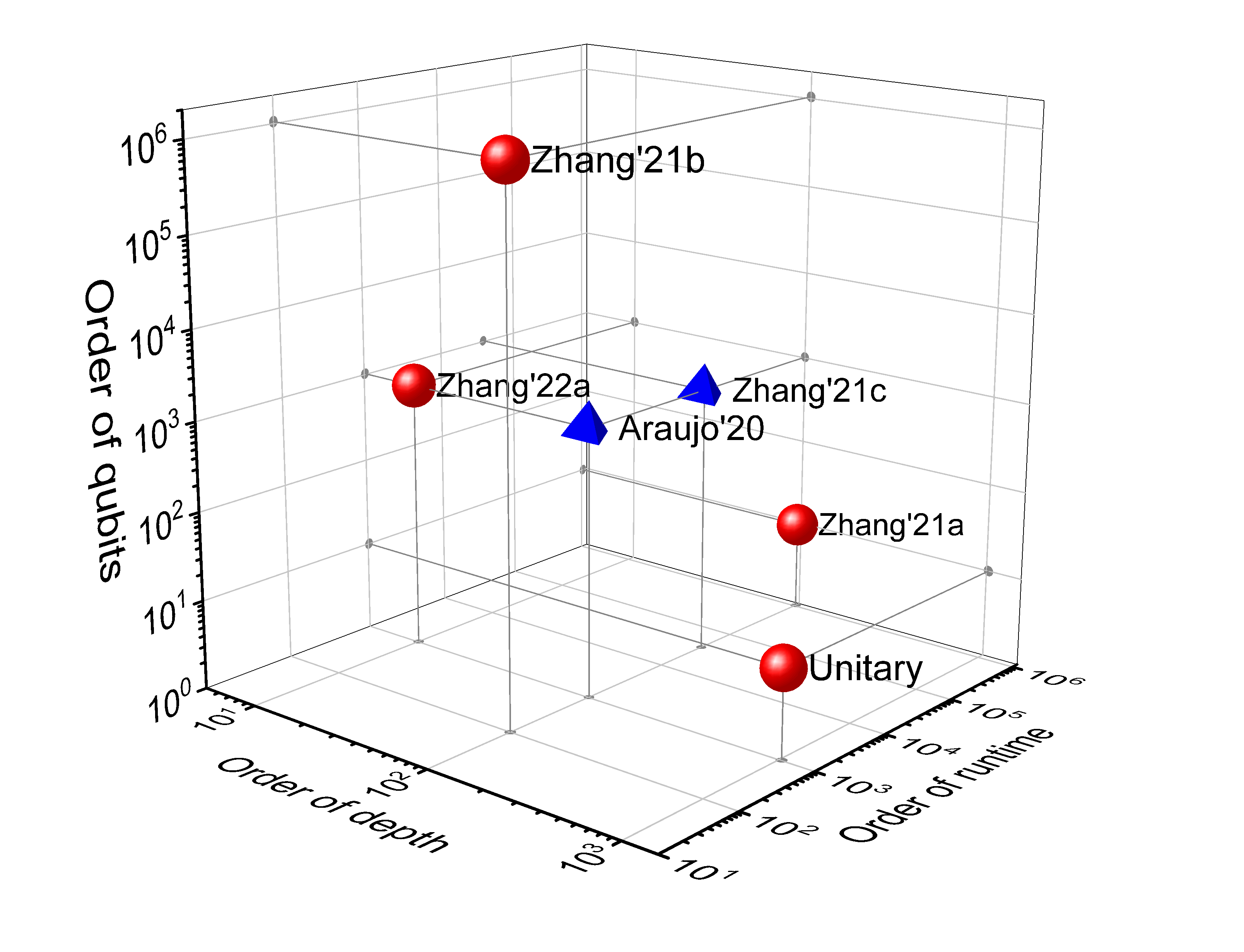}
        \caption{$n=10$}
        \label{fig:sub1}
    \end{subfigure}
    \begin{subfigure}[b]{0.32\textwidth}
        \centering
        \includegraphics[width=\textwidth]{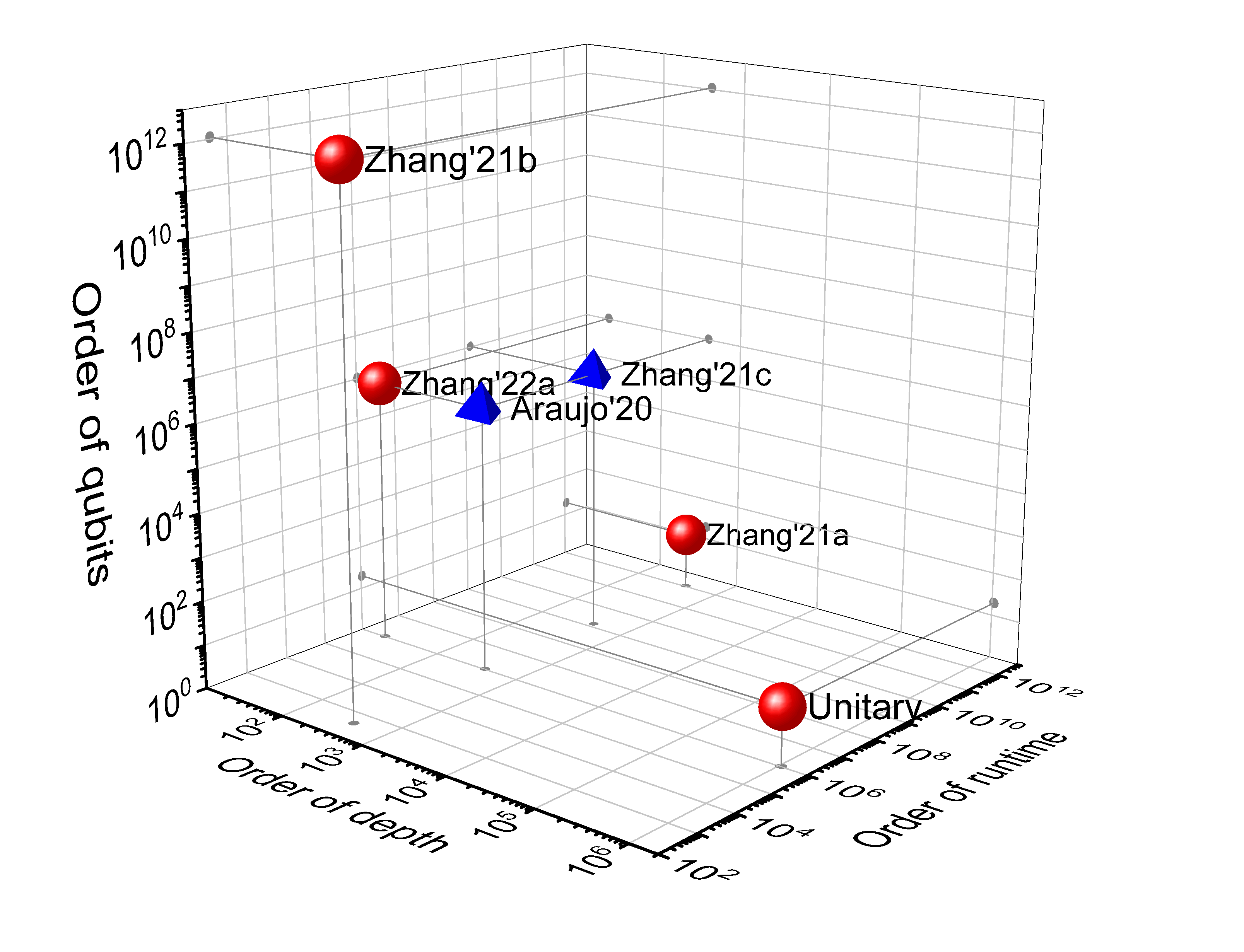}
        \caption{$n=20$}
        \label{fig:sub2}
    \end{subfigure}
    \begin{subfigure}[b]{0.32\textwidth}
        \centering
        \includegraphics[width=\textwidth]{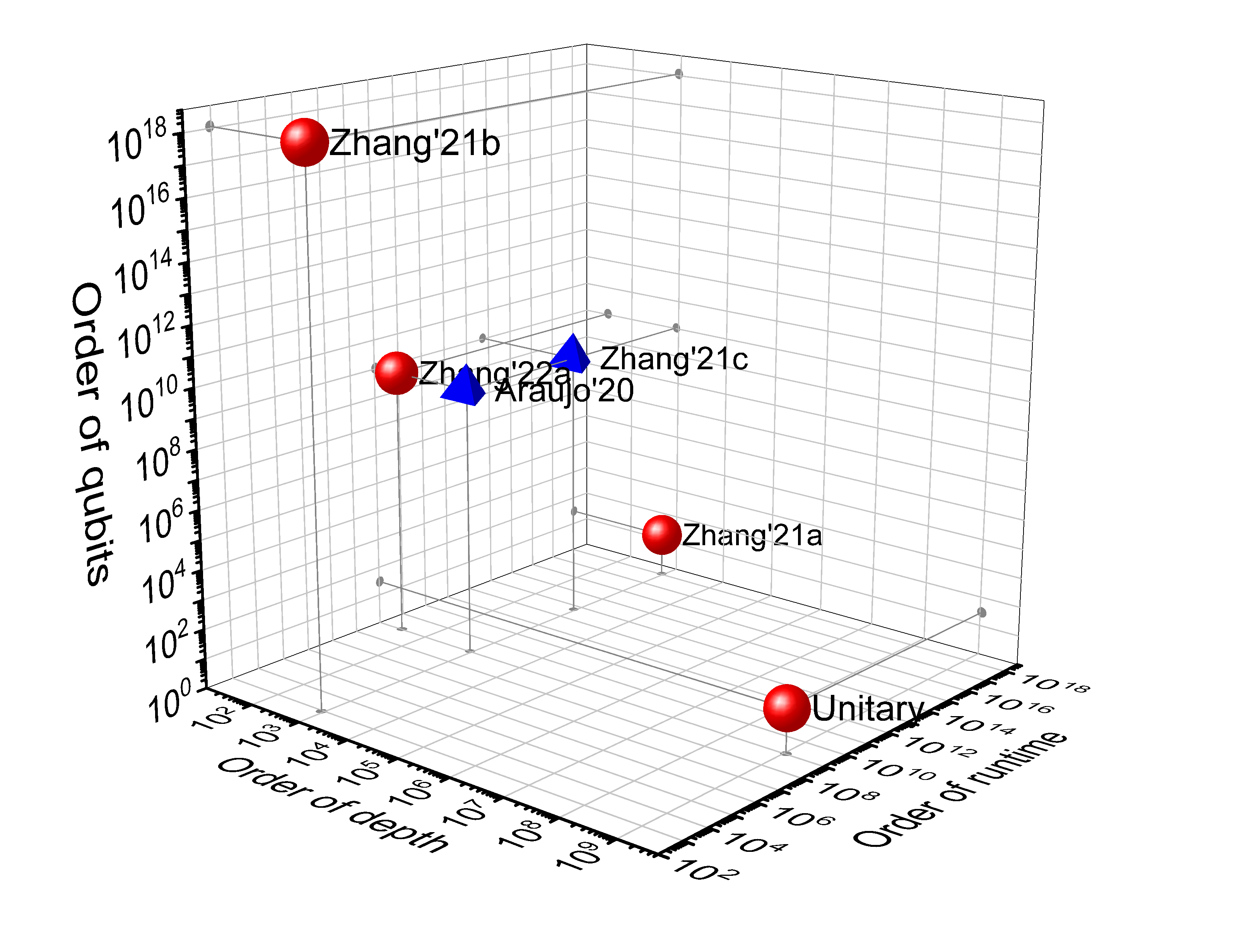}
        \caption{$n=30$}
        \label{fig:sub3}
    \end{subfigure}
    \caption{Relation between the order of circuit depth (labelled ``order of depth''), classical runtime (labelled ``order of runtime''), and qubit count (labelled ``order of qubits'') for algorithms operating on dense statevector representation. Red spheres represent algorithms in the Pareto set; blue tetrahedra~--- those not in the set. }
    \label{fig:dense_order}
\end{figure*}

To better visualize their relationships, we plot the order of complexity values for $n = 10, 20, 30$ qubits in Figure~\ref{fig:dense_order}. Given that statevectors are dense, we set $N=2^n$.

In the figure, the order of complexity values increase with $n$. There are four groups of algorithms in the Pareto set for all values of $n$: Unitary, Zhang'21a, Zhang'21b, and Zhang'22a. While all four groups are theoretically attractive, in practice, we can reduce this set to Zhang'21a and the family of Unitary algorithms. This reduction is due to the prohibitively large (even for future quantum computers) qubit count required for the other two algorithms (Zhang'21b, and Zhang'22a).

Using this method, we reduced the number of algorithms from six to two and are left with the choice between Unitary and Zhang'21a. The difference is that the Unitary algorithms have a larger order of circuit depths, while Zhang'21a offers more optimized quantum circuitry, but a much higher classical runtime. Thus, we have to trade off computation time on a classical device for computation time on a quantum device. For today’s noisy quantum computers, a higher classical runtime is typically better than a higher quantum runtime (as the deeper the circuit, the greater the noise).

This shows how we can reduce the choice of algorithms from six groups to two. In summary, there is no ``free lunch'' for any of the algorithms that convert dense statevectors. As mentioned in Section~\ref{sec:intro}, statevector convertion is an exponentially complex problem. Algorithms must adhere to this boundary, trading off which quality attributes will grow exponentially. This illustrates the difficulty of statevector conversion and, hence, the challenge of loading classical data into the quantum device.

\subsection{Sparse statevector representation}\label{sec:sparse}
Table~\ref{tbl:complexity} gives the order of complexity for four groups\footnote{Note that the NR-group group contains three algorithms. Two of them~\cite{deVeras2020circuit,khan2022ep} extend Probabilistic Quantum Memories~\cite{pqmct, trugenbergercarlofull} and one~\cite{malvetti2021quantum} uses Householder decompositions~\cite{householder1958unitary}.
Technically, the classical runtime for~\cite{malvetti2021quantum} is $O\left(\binom{n}{\ceil{\log_2 \text{nnz}(r)}} + nr\right)$, where $\text{nnz}(r)$ is the count of non-zero bits in an data point and the binomial coefficient represents the number of splitting attempts. The authors of~\cite{malvetti2021quantum} suggest limiting the number of attempts to some constant value, reducing the complexity to $O(nr)$.
} of algorithms (six algorithms in total) operating on sparse statevectors. As in Section~\ref{sec:dense}, the algorithms are grouped by distinct values of the order of circuit depth, classical runtime, and qubit count.

As we can see from the order of complexity formulas in Table~\ref{tbl:complexity}, these algorithms cannot deal efficiently with dense statevectors (as $r \to 2^n$). Nevertheless, they are substantially more efficient in sparse cases (where $r \ll 2^n$). 

\begin{figure*}[tb]
    \centering
    \begin{tabular}{ccc}
        \begin{subfigure}[b]{0.32\textwidth}
            \centering
            \includegraphics[width=\textwidth]{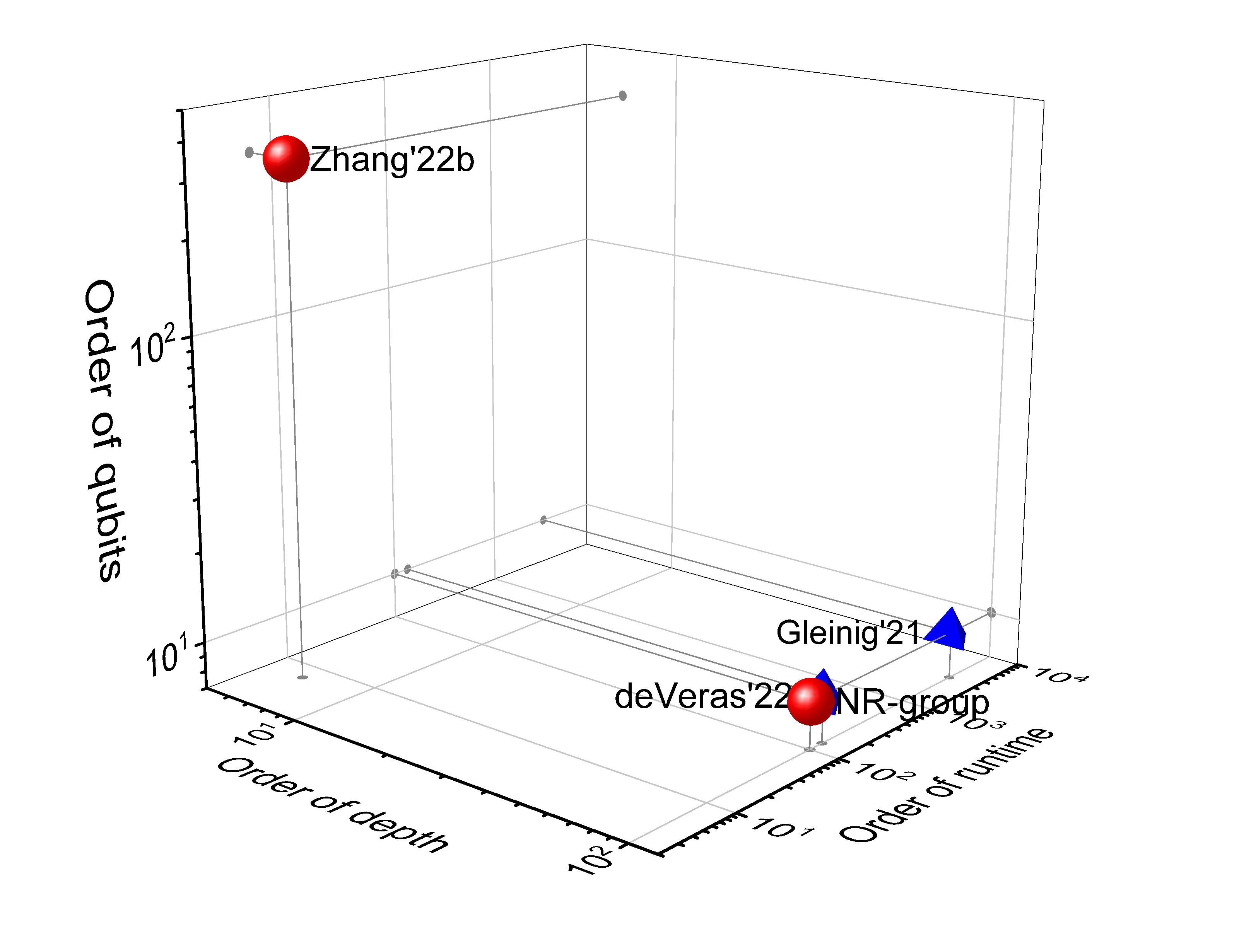}
            \caption{$n=10, r=10$}
            \label{fig:subfig1}
        \end{subfigure} &
        \begin{subfigure}[b]{0.32\textwidth}
            \centering
            \includegraphics[width=\textwidth]{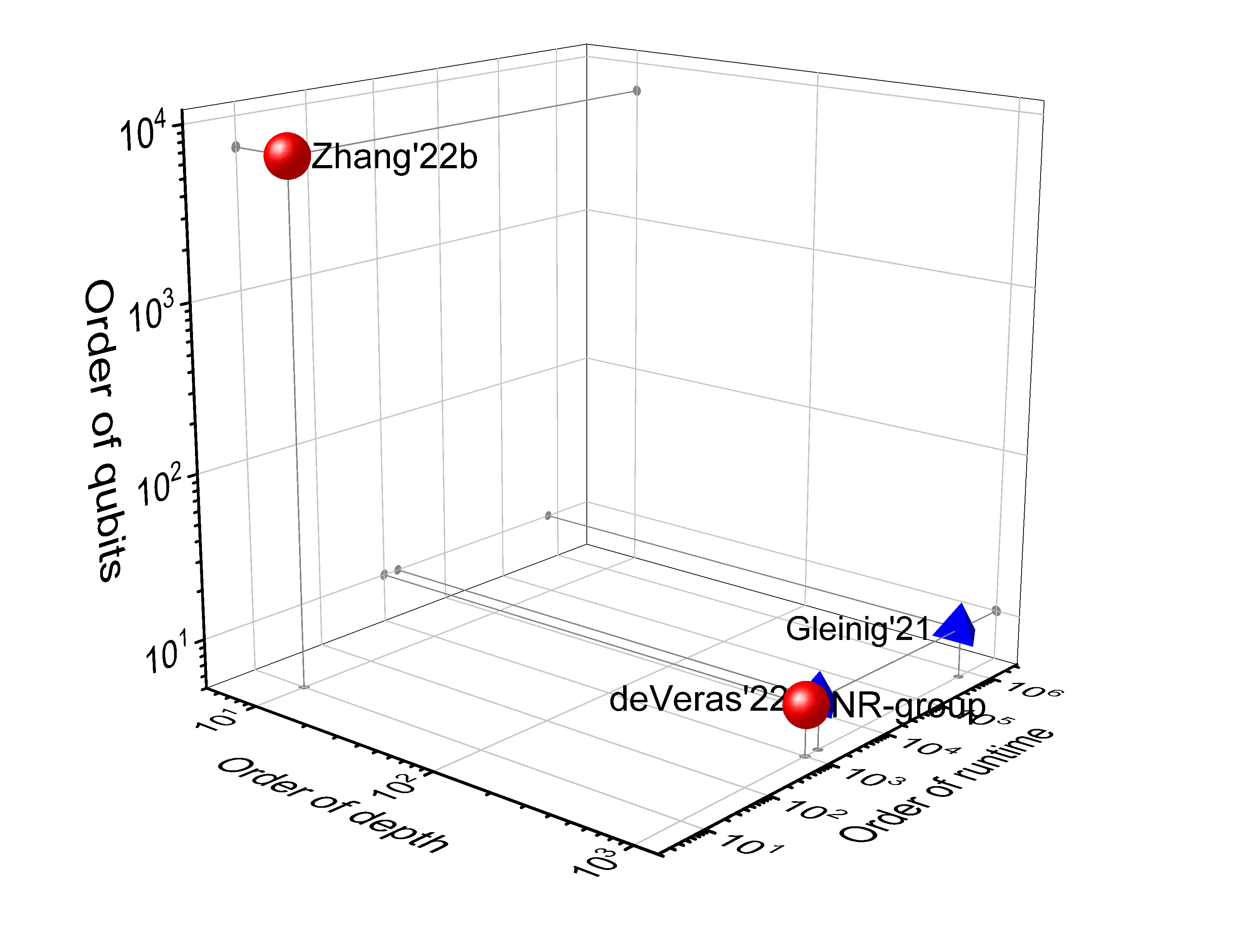}
            \caption{$n=10, r=100$}
            \label{fig:subfig2}
        \end{subfigure} &
        \begin{subfigure}[b]{0.32\textwidth}
            \centering
            \includegraphics[width=\textwidth]{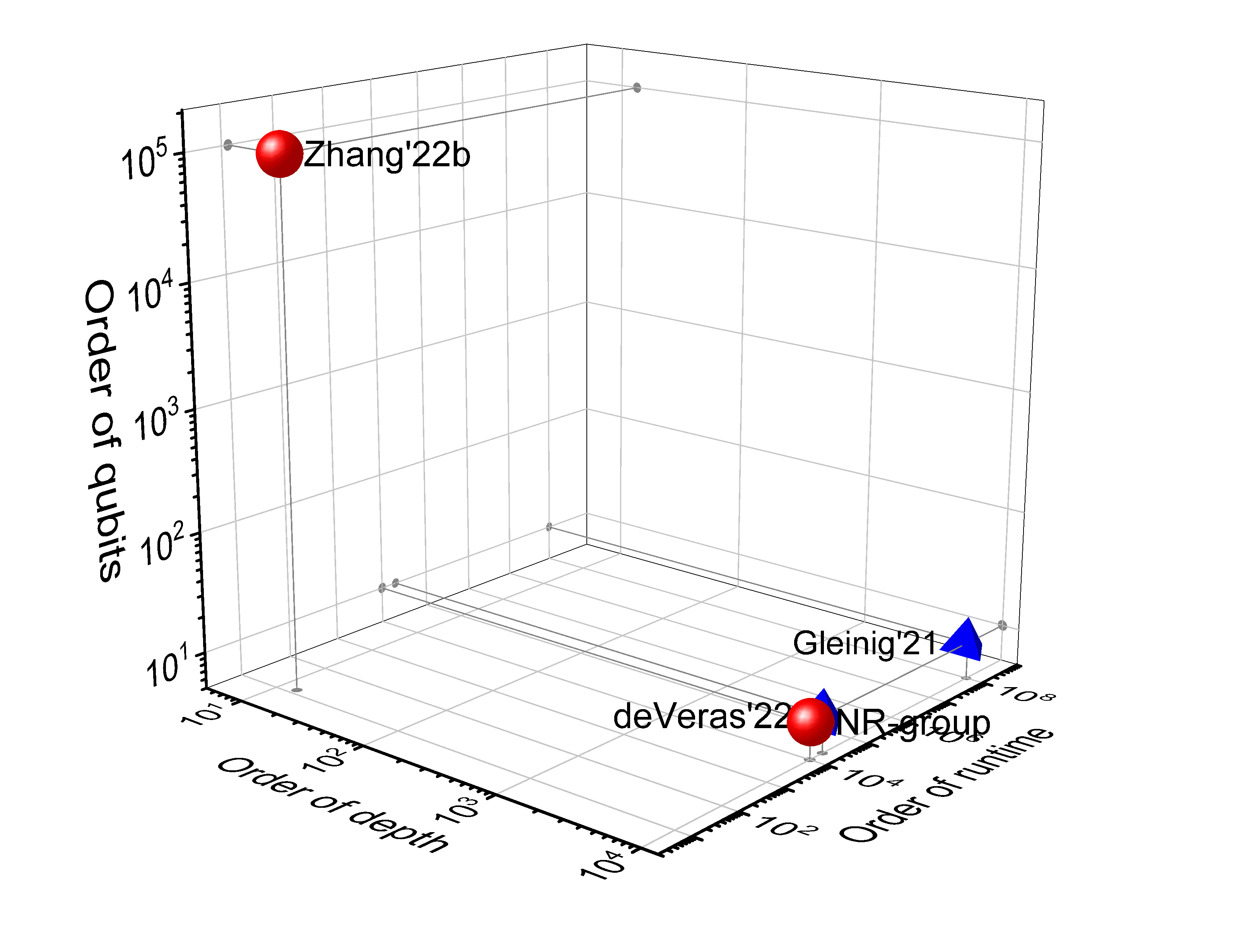}
            \caption{$n=10, r=1000$}
            \label{fig:subfig3}
        \end{subfigure} \\
        \begin{subfigure}[b]{0.32\textwidth}
            \centering
            \includegraphics[width=\textwidth]{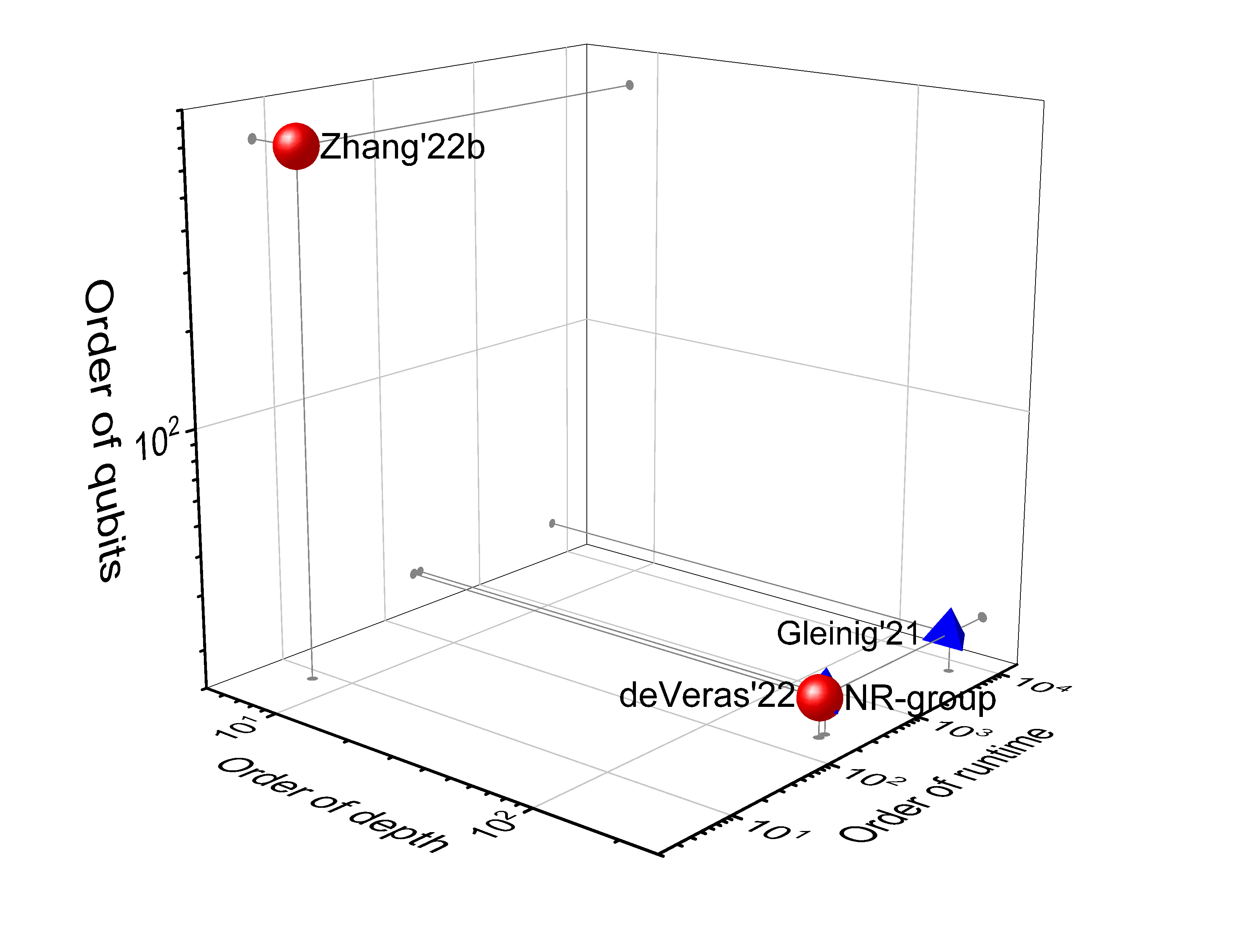}
            \caption{$n=20, r=10$}
            \label{fig:subfig4}
        \end{subfigure} &
        \begin{subfigure}[b]{0.32\textwidth}
            \centering
            \includegraphics[width=\textwidth]{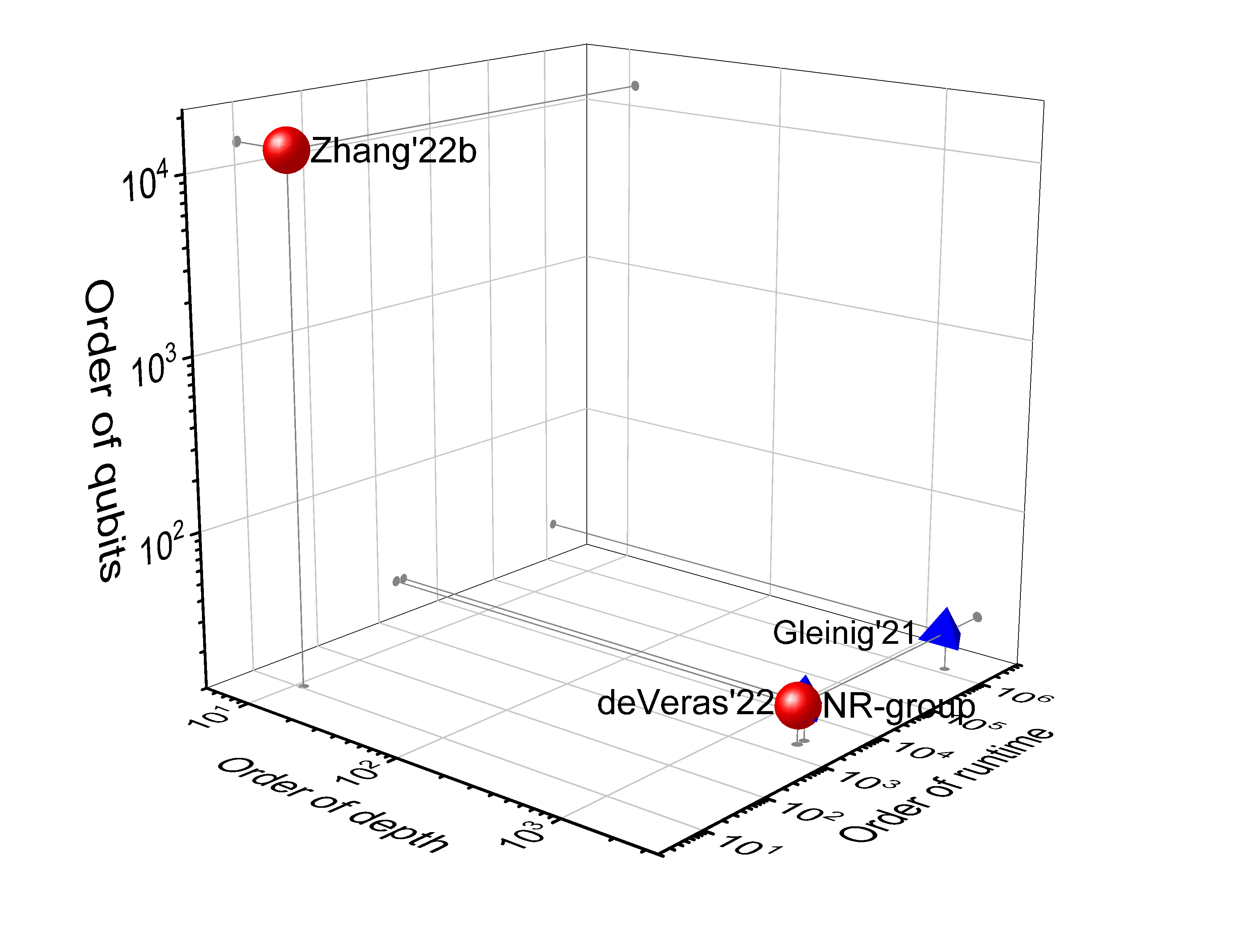}
            \caption{$n=20, r=100$}
            \label{fig:subfig5}
        \end{subfigure} &
        \begin{subfigure}[b]{0.32\textwidth}
            \centering
            \includegraphics[width=\textwidth]{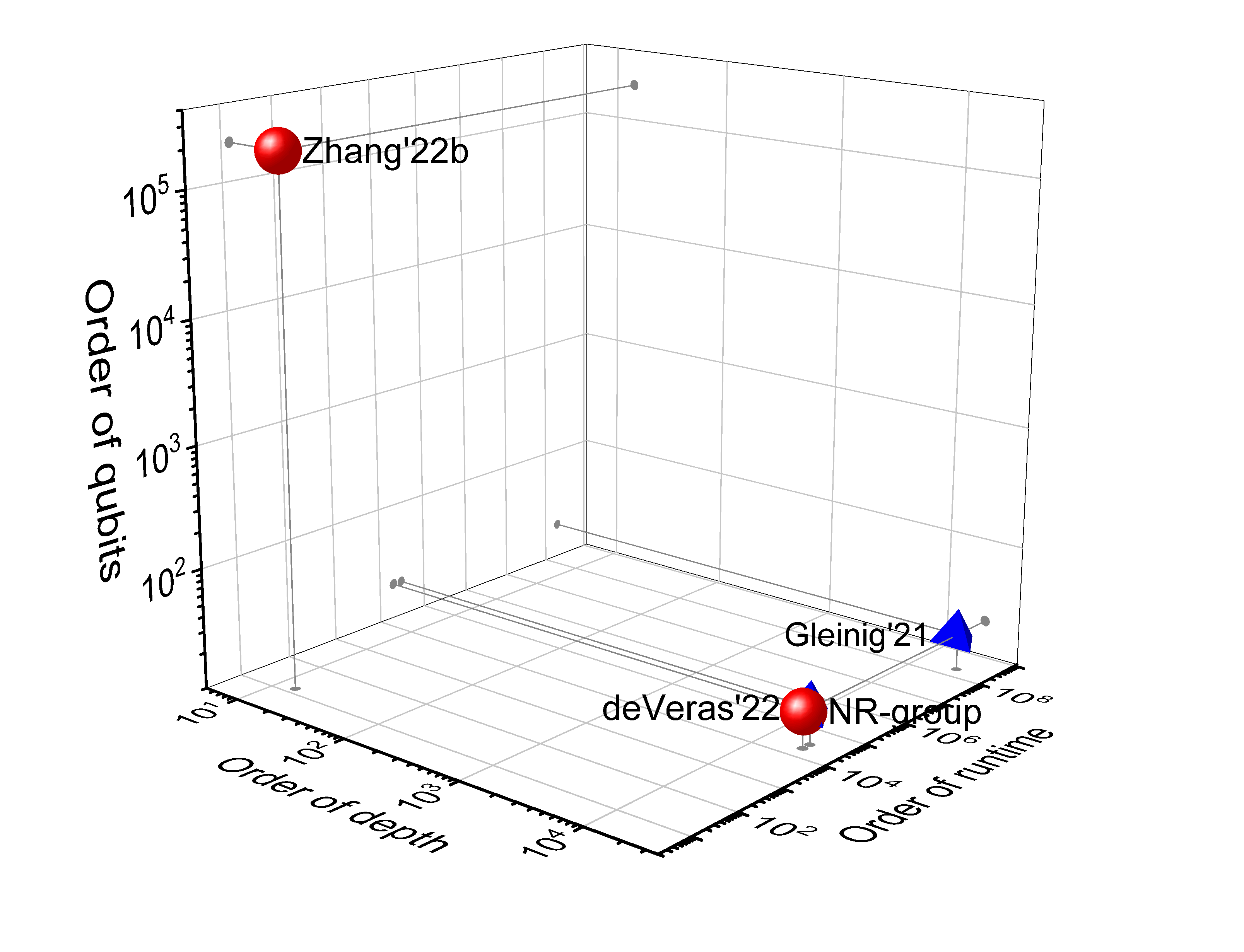}
            \caption{$n=20, r=1000$}
            \label{fig:subfig6}
        \end{subfigure} \\
        \begin{subfigure}[b]{0.32\textwidth}
            \centering
            \includegraphics[width=\textwidth]{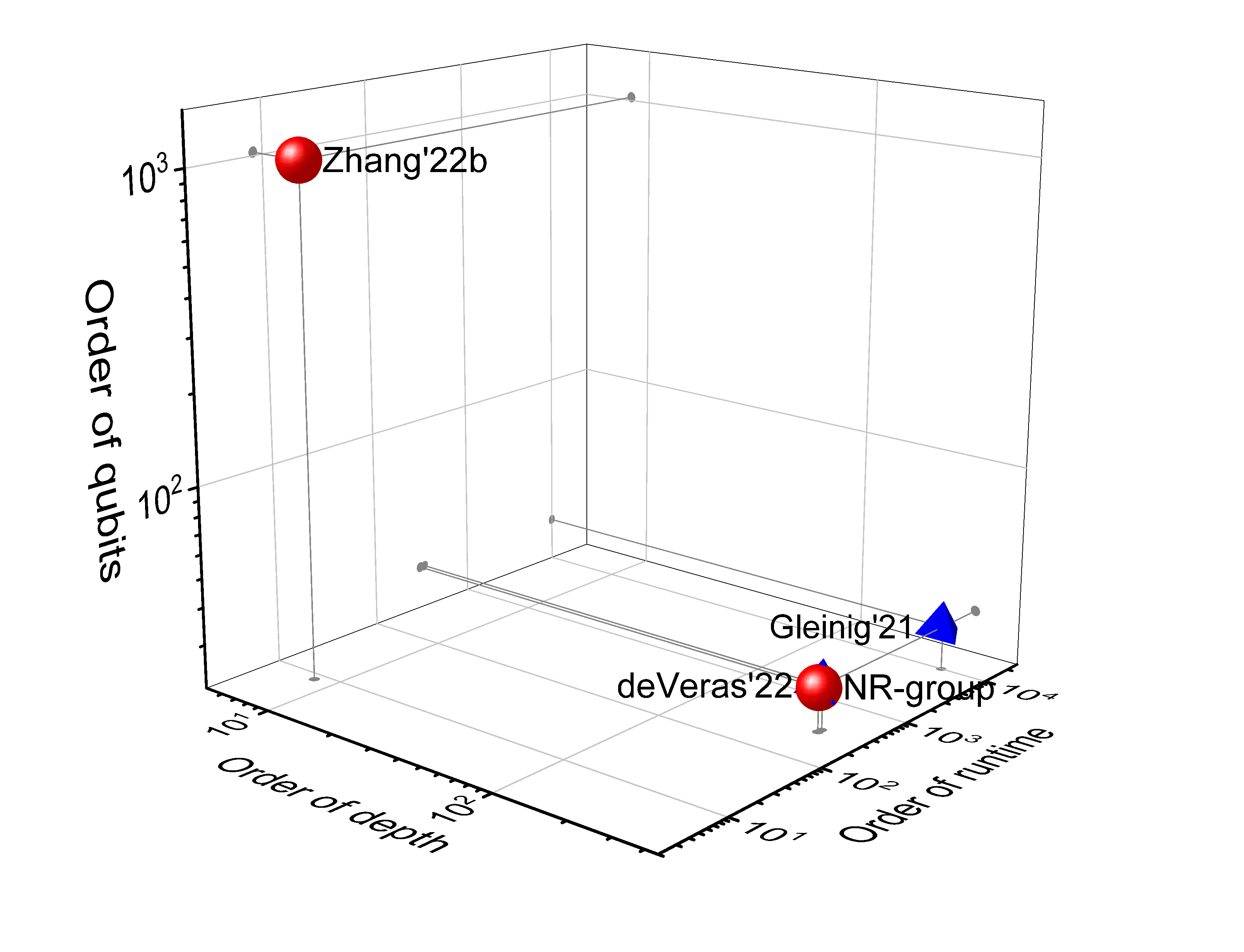}
            \caption{$n=30, r=10$}
            \label{fig:subfig7}
        \end{subfigure} &
        \begin{subfigure}[b]{0.32\textwidth}
            \centering
            \includegraphics[width=\textwidth]{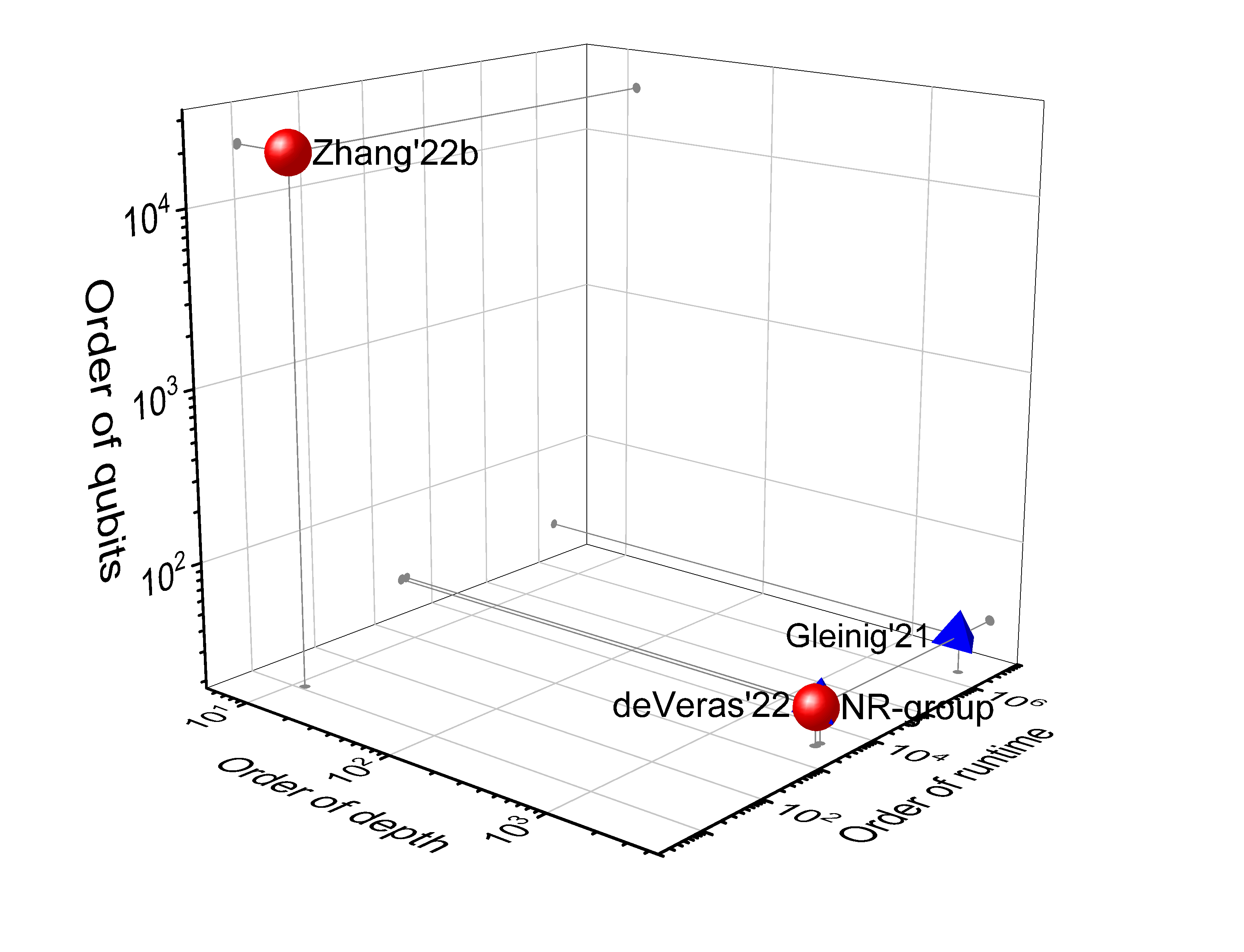}
            \caption{$n=30, r=100$}
            \label{fig:subfig8}
        \end{subfigure} &
        \begin{subfigure}[b]{0.32\textwidth}
            \centering
            \includegraphics[width=\textwidth]{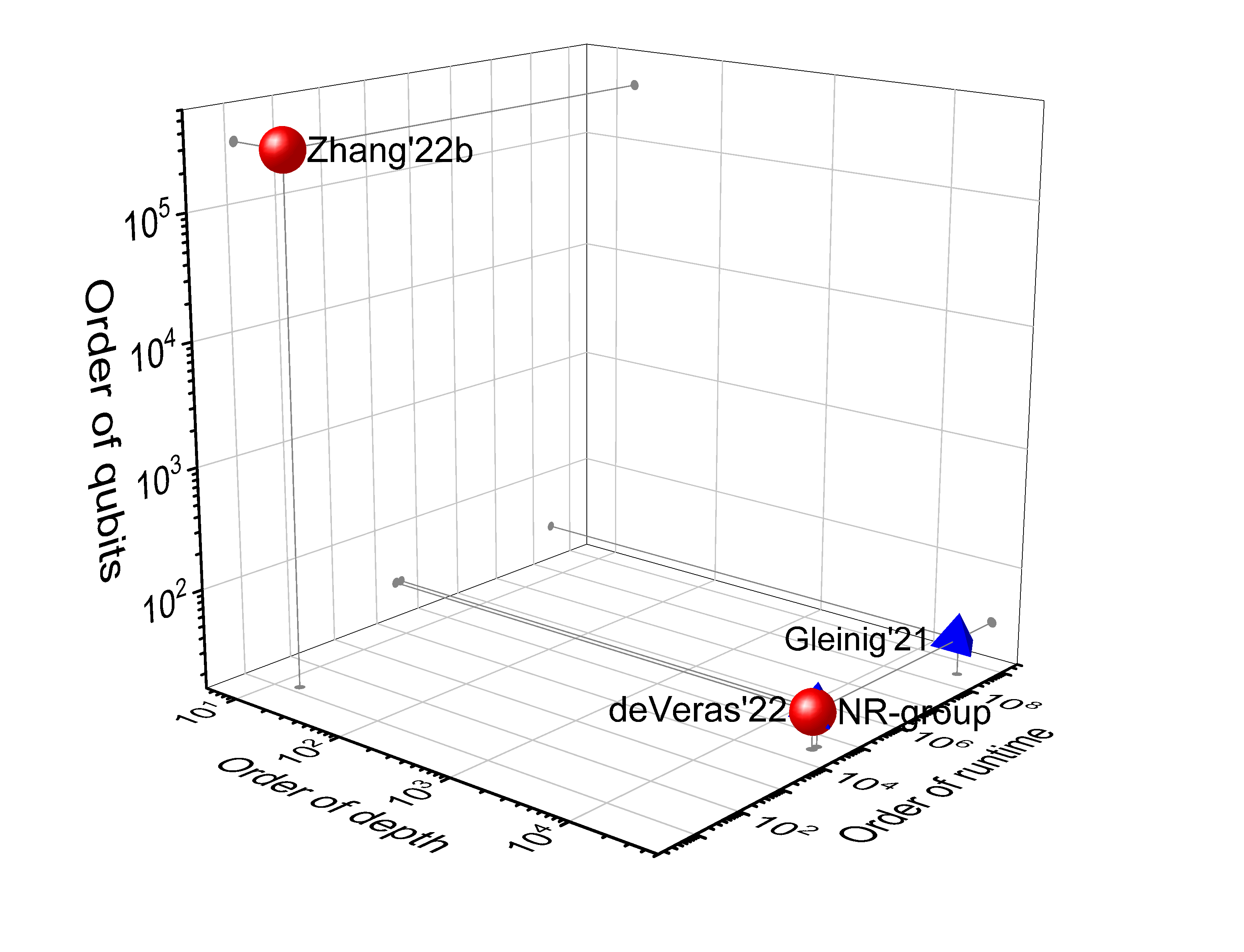}
            \caption{$n=30, r=1000$}
            \label{fig:subfig9}
        \end{subfigure}
    \end{tabular}
        \caption{Relation between the order of circuit depth (labelled ``order of depth''), classical runtime (labelled ``order of runtime''), and qubit count (labelled ``order of qubits'') for algorithms operating on sparse statevector representation. We demonstrate various combinations of $n$ and $r$ to illustrate how fast growth is with these two factors. Red spheres represent algorithms in the Pareto set; blue tetrahedra~--- those not in the set. }
    \label{fig:sparse_order}
\end{figure*}

To visualize the relationships of quality attiibutes, we plot the order of complexity values for the Cartesian product $n = 10, 20, 30$ qubits and $r = 10, 100, 1000$ non-zero elements in Figure~\ref{fig:sparse_order}. Some algorithms report complexity based on the number of non-zero bits in an element. We choose the worst-case scenario for all $n$ bits equal to $1$ to simplify the comparison.

For all values of $n$ and $r$, there are two groups of algorithms in the Pareto set: NR-group and Zhang'22b. NR-group requires fewer qubits, but produces a deeper circuit using a significant amount of classical runtime; Zhang'22b produces a shallow circuit quickly, but requires a large number of qubits. Even though Zhang'22b approach is impractical today, it has great promise for future quantum computers with many qubits.

By ranking algorithms, we focused on two of four groups. One can choose from three algorithms in the NR-group subset if one needs an algorithm for a modern quantum computer\footnote{At the time of writing, high levels of gate errors and decoherence make deep circuits impractical until error correction schemes (such as~\cite{bravyi2024high}) are introduced. IBM, for example, hopes to provide such computers by 2029~\cite{ibmroadmap}.} with fewer qubits. 
As an alternative, Zhang'22b would be a good choice for quantum computers with large numbers of qubits in the future.

In summary, sparse state vector algorithms require fewer resources (as long as $r \ll 2^n$) than dense state vector algorithms, which makes them promising for practical applications. However, we should still wait until quantum computers mature (decoherence goes down and the qubit count goes up) before we can load relatively large volumes of classical data.

\subsection{Alterable circuits}\label{sec:alter}
Assessing the presence of the fifth feature, the ability to alter circuits, is challenging because no author explicitly addresses it. We conjecture that at least three algorithms can be altered for online tasks~\cite{deVeras2020circuit, deVeras2022double, khan2022ep} (since they are based on Probabilistic Quantum Memories~\cite{pqmct, trugenbergercarlofull} and FF-QRAM~\cite{park2019circuit}). These algorithms process data points independently and do not require post-processing, making identifying and eliminating the gates associated with a particular data point simple.

We are uncertain whether altering the other algorithms is possible or easy. This uncertainty presents an interesting avenue for future research.

\section{A note on multi-qubit gate count}
In our assessment of quality attributes, we emphasize the depth of the circuit without differentiating between single- and multi-qubit gates. 

In practice,  multi-qubit gates can be decomposed into several two-qubit \textsc{cnot} gates and single-qubit gates. Thus, researchers often report the count of \textsc{cnot} gates for a given algorithm.

Two-qubit gates are computationally more expensive (moreover, they often require additional expensive \textsc{swap} gates for modern architectures) and introduce more noise than single-qubit gates.  As a result, \textsc{cnot} gate count is sometimes beneficial as an additional quality metric.

Often, circuit depth and \textsc{cnot} gate count correlate. For example, for algorithms operating on sparse statevectors, discussed in this paper, the order of complexity of circuit depth and the order of complexity of \textsc{cnot} count are the same.

Thus, readers may consider including the count of two-qubit gates as an additional quality attribute.

\section{Summary}

In this study, we identify and define five quality attributes for statevector preparation algorithms: circuit depth, qubit count, classical runtime, statevector representation (dense or sparse), and the ability to dynamically alter the circuit.

We propose a methodology for comparing various state preparation algorithms using multi-objective optimization techniques. This methodology involves reducing the set of algorithms based on quality attributes and computing a Pareto set to determine which algorithms offer the most desirable combined properties (other multi-objective techniques can also be used). A visual representation illustrates the relationship between circuit depth, classical runtime, and qubit count. These visualizations help understand trade-offs and choose appropriate algorithms based on specific needs.

With an illustrative example, we compare seven groups of algorithms operating on dense statevectors and six groups~---~on sparse statevectors.

To conclude, our study highlights the complex trade-offs involved in the preparation of quantum states from classical data and proposes a structured approach for evaluating and selecting the most suitable algorithms. By balancing different quality attributes, we aim to guide researchers and practitioners in optimizing state preparation for various quantum computing tasks.

\bibliographystyle{IEEEtran}
\bibliography{references}

\end{document}